\documentclass[journal]{IEEEtran}
\usepackage{subfigure}
\usepackage{algpseudocode}
\usepackage{algorithm}
\usepackage{cite}

\ifCLASSINFOpdf
   \usepackage[pdftex]{graphicx}

\else

\fi

\usepackage[cmex10]{amsmath}
\usepackage{url}

\usepackage{amssymb}

\usepackage{epstopdf}

\begin{document}
\title{Distributed Cross-layer Dynamic Route Selection in Wireless Multiuser Multihop Networks}

\author{Kamal~Rahimi Malekshan,~\IEEEmembership{Student \!Member IEEE,}
       and~Farshad~Lahouti,~\IEEEmembership{Senior \!Member IEEE}% <-this % stops a space
\thanks{Kamal Rahimi Malekshan is with the Department of Electrical and Computer Engineering, University of Waterloo, Canada. Email: krahimim@uwaterloo.ca.

Farshad Lahouti is with the Center for Wireless Multimedia Communications, Center of Excellence in Applied Electromagnetic Systems, School of Electrical and Computer Engineering, College of Engineering, University of Tehran, Tehran, Iran. Email: lahouti@ut.ac.ir.
}
}

%\markboth{Journal of \LaTeX\ Class Files,~Vol.~6, No.~1, January~2007}%
%{Shell \MakeLowercase{\textit{et al.}}: Bare Demo of IEEEtran.cls for Journals}

\maketitle

\begin{abstract}
%\boldmath
In wireless ad-hoc networks, forwarding data through intermediate relays extends the coverage area and enhances the network throughput. We consider a general wireless multiuser multihop transmission, where each data flow is subject to a constraint on the end-to-end buffering delay and the associated packet drop rate as a quality of service (QoS) requirement. The objective is to maximize the weighted sum-rate between source-destination pairs, while the corresponding QoS requirements are satisfied. We introduce two new distributed cross-layer dynamic route selection schemes in this setting that are designed involving physical, MAC, and network layers. In the proposed opportunistic cross-layer dynamic route selection scheme, routes are assigned dynamically based on the state of network nodes' buffers and the instantaneous state of fading channels. In the same setting, the proposed time division cross-layer dynamic route selection scheme utilizes the average quality of channels instead for more efficient implementation. Detailed results and comparisons are provided, which demonstrate the superior performance of the proposed cross-layer dynamic route selection schemes.\\
\end{abstract}


% \fi

\IEEEpeerreviewmaketitle

\section{Introduction}

\IEEEPARstart{M}{obile} ad-hoc networks (MANETs) are formed by a number of mobile nodes communicating with each other over wireless channels without a pre-established infrastructure. These nodes can dynamically organize arbitrary ad-hoc network topologies. As the transmission range of devices are often limited, the packets may need to be forwarded by one or more intermediate relays before they reach their destinations. This paper focuses on designing distributed route selection strategies that aim at maximizing the quality of service (QoS) constrained network throughput, while taking into account both buffers and channels conditions in a cross-layer manner.\\

Research from network and information theory perspectives has led to a number of relaying schemes in MANETs \cite{Gastpar, Draves, Laneman, Wang}. In \cite{Laneman} and \cite{Wang}, the authors consider spectral efficiency as a performance measure to select a desired route from one source to a destination in wireless networks without taking into account the effect of other users requesting service in the network. They assume that data is transmitted by the source and forwarded by the intermediate relays without a store and wait stage. Other transmission schemes are proposed in \cite{Goldsmith2,Goldsmith3,Mardani} in which the nodes store data in their buffers, and employ adaptive modulation and coding (AMC) and automatic repeat request (ARQ) to enhance network throughput. The trade-off between delay, diversity and multiplexing in a multihop wireless network with ARQ is studied in \cite{Goldsmith2}. The problem of finding the path with the minimum end-to-end outage probability from one source to a destination is studied in \cite{Babaee}.\\

The selection of a route and corresponding relays for transmission between one specific source and destination pair, may affect the communication of other nodes. Hence, the existence of other source destination pairs in the network and the varying nature of wireless networks indicate that the routes and relays are to be assigned dynamically to guarantee the QoS requirements. Multipath routing decreases the effect of unreliable wireless links in a constantly changing topology and improves QoS of MANETs \cite{Nasipuri,Harms,Mueller, Dhaka,Ai}. In fact, starting from single path routing, significant performance gains are achieved by employing a limited number of additional paths, beyond which the potential gain diminishes \cite{Nasipuri}. In addition, providing alternative paths to intermediate nodes is more advantageous than doing that only to the source node \cite{Nasipuri}. Multipath routing can also help distribute the transmission load more evenly, which is vital to protect wireless nodes from power depletion \cite{Harms}. Multipath transmission from one source node to multiple destinations with the aid of multiple relays is considered in \cite{Melo}. At each instant, the source node chooses a destination for transmission and selects an intermediate relay node to forward the packet. The destination receives both signals transmitted by the source and relay, and uses a selective combining strategy to decode the signal. The authors in \cite{Ronasi} applied a combination of multipath routing and adaptive channel coding to improve delay and throughput performance of a wireless network. In \cite{Tsirigos}, a multipath routing scheme is proposed for transmission over unreliable wireless ad-hoc links that increases the probability of successful reception for the essential part of information at the destination node.\\

In this paper, we propose two new cross-layer dynamic route selection schemes for a multiuser multihop ad-hoc network. The proposed schemes distributedly direct information between source destination pairs through varying scheduled routes. As we demonstrate, the proposed schemes enhance the overall throughput of the network and help meet the QoS requirements. The optimized designs in fact involve physical, MAC, and network layers. To the best of our knowledge, this is the first work which considers cross-layer resource allocation across the three layers in a general multihop ad-hoc network. The opportunistic cross-layer dynamic route selection (OCDR) efficiently and dynamically assigns routes to source-destination pairs based on the state of network nodes' buffers and the instantaneous condition of fading channels. The proposed time division cross-layer dynamic route selection (TCDR) schedules the assignment of varying routes within the same framework, but utilizing the average link qualities instead. In both schemes, the nodes exchange their average service rate information with one hop neighbors, which also reflect their buffers conditions. The nodes then identify the optimum dynamic route selection plan for forwarding data toward destinations in a distributed manner.\\

The remainder of this paper is organized as follows. We introduce the system model in Section II. Section III presents our proposed dynamic route selection schemes. The simulation results are presented in Section IV. Finally, Section V concludes the paper.\\

\begin{figure}
\centering
\includegraphics[width=4.1in, trim=1.5cm 8.5cm 1cm 6.5cm, clip=true]{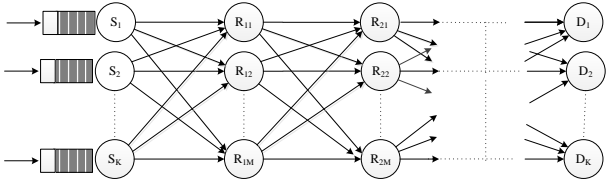}
\caption{System model for a multihop transmission in a network consisting of $K$ source-destination pairs.} \label{Network}
\end{figure}

\section{Preliminaries}
Consider multiuser multihop transmission in a network consisting of $K$ source-destination pairs that communicate through intermediate relays. The equivalent two-hop model is considered in \cite{kamal0}. We consider $L$ hop transmission with $M$ relaying nodes at each hop. The source, destination, and relays at hop $l\in\{1,2,...,L\}$ are identified by ${\cal S}=\{S_1,\hdots,S_K\}$, ${\cal D}=\{D_1,\hdots,D_K\}$, and ${\cal R}_l=\{R_{l1},\hdots,R_{lM}\}$, respectively. The source nodes forward their data to relays in the first hop. On the path from source to destination, the relays in hop $l\in\{2,3,...,L-1\}$ receive data from relays in hop $l-1$, store data of each source in a distinct buffer, and forward the data to relays in hop $l+1$. The relays in the last hop ($L$) deliver the data packets of each source node to its destination node. A decode and forward protocol is assumed at the relays. The objective is to maximize the weighted sum-rate of communications between source destination pairs. The QoS requirement is quantified by a probabilistic end-to-end delay constraint and its associated packet loss rate. Figure \ref{Network} shows the system model, in which the buffers at the relays are omitted for brevity. For simplified presentation, we define ${\cal R}_i^j\triangleq{\cal R}_i\cup\hdots\cup{\cal R}_j, i,j\in \{1,\hdots,L\}, i<j$.\\

\subsection{Channel Model}
The channels are assumed non-selective, block Rayleigh fading and signals also suffer path loss. Channel time is divided into time slots of duration $T$, where channel gain during a time slot is constant but varies independently between two consequent time slots. At each time slot, the channel state information (CSI) to every next hop neighboring node is perfectly estimated, and reliably fed back to the node without delay. Each node has access to independent bandwidth $W$ in which it transmits with fixed power $P$. It is assumed that given the channel SNR in time slot $t$, $\gamma(t)$, the AMC scheme can achieve the instantaneous capacity of the link. Let $\sqrt{h_{ij}(t)}$ denote the channel gain of the link between nodes $i$ and $j$ at time slot $t$. The capacity between nodes $i$ and $j$ at time slot $t$ is given by
\begin{equation}\label{1}
  C_{ij}(t)=W \log_2\left(1+\frac{P\, h_{ij}(t)}{N_0\, W}\right),
\end{equation}
where $N_0$ is the gaussian noise power density and $\frac{P\, h_{ij}(t)}{N_0\, W}=\gamma_{ij}(t)$  is the instantaneous SNR of received signal at node $j$ from node $i$ at time slot $t$. For Rayleigh fading, the probability density function (PDF) of $\gamma_{ij}$ is
\begin{equation}\label{2}
  P_{\gamma_{ij}}(\gamma )=\frac{1}{\bar{\gamma}_{ij}} e^{-\frac{\gamma}{\bar{\gamma}_{ij}}},
\end{equation}
where $\bar{\gamma}_{ij}$ is the average received SNR of the link between nodes $i$ and $j$.

\subsection{Queuing Network Model}
The packets arrive at the source node $k\in \mathcal{S}$ as a Poisson process with average $\rho_k$. We assume that the random service time process in each node, which depends on the state of Rayleigh fading channel, as an approximation to be i.i.d exponential \cite{Goldsmith2}. This approximation makes the multiuser input, multiuser output, multihop network problem tractable. Packets depart from a node and arrive at the next hop node as Poisson processes \cite{Bolch}. Thus, we can use an $M/M/1$ queue to model packet delay of each source node at source and any of the intermediate relays. Let $\mu_{ik}$ denote the average service rate at node $i$ for data of source node $k$, and $\mu_{ijk}$ be the average service rate at node $i$ to next hop node $j$ for data of source node $k$. We have $\mu_{ik}=\sum_{j} \mu_{ijk}$. Let $\rho_{ik}$ be the average arrival data rate of source $k$ at node $i$ and $\rho_{ijk}$ be the average data rate of source $k$ that node $i$ forwards to next hop node $j$. Note that $\rho_k=\rho_{kk}, k\in \mathcal{S}$ is the average arrival data rate at source $k$, and $\rho_{ik}=\sum_j \rho_{ijk}$. If the packet loss rate of source $k$ at node $i$ is small, we have
\begin{equation}\label{5}
  \rho_{ijk}=\frac{\mu_{ijk}}{\mu_{ik}}\times\rho_{ik}.
\end{equation}
When the queue is stable, i.e., $\rho_{ik}<\mu_{ik}$, the probability that delay at node $i$ for data of source $k$, $D_{ik}$, exceeds a deadline $D_{ik}^*$ is given by \cite{Goldsmith2}
\begin{equation}\label{7}
  \epsilon_{ik}^*\triangleq P(D_{ik}>D_{ik}^*)=\frac{\rho_{ik}}{\mu_{ik}}e^{-D_{ik}^*(\mu_{ik}-\rho_{ik})}.
\end{equation}
If $\epsilon_{ik}^*\ll \mu_{ik}/\rho_{ik}$, we have
\begin{equation}\label{8}
  D_{ik}^*=\frac{\ln (\frac{\rho_{ik}}{\epsilon_{ik}^*\mu_{ik}})}{\mu_{ik}-\rho_{ik}}\thickapprox \frac{\ln(\frac{1}{\epsilon_{ik}^*})}{\mu_{ik}-\rho_{ik}}.
\end{equation}

\subsection{Problem Formulation}
As a QoS requirement, we consider a limited probability of violating the delay bound. In other words, the probability that end-to-end delay of source $k$ packets exceed $D_k^*$ is to be smaller than a threshold. To achieve this goal, we can limit the delay of source $k$ at each node $i$ (on the path from source to destination) to $D_{ik}^*$, such that $\sum_{i\in{path}} D_{ik}^* \leq D_{k}^*$ \cite{Goldsmith3}. Thus, the packets of source $k$ are removed from the queue in node $i$ if they are not transmitted before the deadline $D_{ik}^*$. Similarly, to achieve the end-to-end packet loss rate $\epsilon_{k}^*$ for data of source $k$, we restrict packet loss rate due to delay violation at node $i$ (on the path from source to destination) to $\epsilon_{ik}^*$, such that $1-\prod_{i\in path}(1-\epsilon_{ik}^*)\leq \epsilon_{k}^*$.\\

The objective in this work is to maximize the weighted sum arrival rate at the source nodes, while the end-to-end delay violation probability (and its associated packet drop rate) is constrained. This design problem may be formulated as follows
\begin{align}
  \max F&=\sum_{k\in \mathcal{S}} f_k\rho_k \label{9}\\
  \text{s.t.: }& P(D_{ik}>D_{ik}^*)<\epsilon_{ik}^*,\label{91}\\ k\in&\mathcal{S},i\in\mathcal{S}\cup\mathcal{R}_1^L,\nonumber
\end{align}
where $f_k$ is the assigned weight to data rate of source $k$.\\

\section{Cross-Layer Dynamic Route Selection}
In cross-layer dynamic route selection data is transmitted between source-destination pairs through dynamic routes. This is in contrast to static routing where a fixed multihop link is established between source-destination pairs. The proposed cross-layer dynamic route selection schemes at the MAC layer schedule the assignment of relays in different hops based on the associated channel and buffer conditions. As we shall demonstrate, the proposed schemes enhance the overall throughput of the network and help meet the diverse QoS requirements.\\

\subsection{Opportunistic Cross-layer Dynamic Route Selection}
In the proposed opportunistic cross-layer dynamic route selection (OCDR), the path for data transmission is dynamically adjusted based on the instantaneous condition of fading channels and the traffic loads of network nodes. In each time slot, each node transmits over the link with the highest instantaneous weighted SNR in comparison with other possible links for transmission. The assigned weights for the SNR of each link are adjusted to meet the QoS requirements of users in the network.\\

Let $U_i$ denote the set of next hop nodes for node $i\in{\mathcal{S},\mathcal{R}_1^{L}}$. At time slot $t$, node $i$ chooses to transmit to the next hop node ${\hat j}\in U_i$, if
\begin{equation}\label{10}
  {\hat j}=\arg\max_{j\in U_i} \Phi_{ij}(t),
\end{equation}
where $\Phi_{ij}(t)$ is the priority function in node $i$ for transmission to next hop node $j\in U_i$ at time slot $t$. If multiple links have the same value of $\Phi_{ij}(t)$, one of them is selected randomly. We define
\begin{equation}\label{11}
  \Phi_{ij} (t)\triangleq \beta_{ij}\frac{\gamma_{ij}(t)}{\bar{\gamma}_{ij}}
\end{equation}
where $\beta_{ij}$, are set as described below. While next hop node $j$ is selected for transmission at node $i$, the allocated fraction of time to transmit data of source $k\in \mathcal{S}$ is $\alpha_{ijk}$. Thus, when next hop node $j$ is selected for transmission at node $i$, data of source node $k$ is transmitted with probability
\begin{equation}\label{11a}
  \pi_{ijk}=\frac{\alpha_{ijk}}{\sum_{k'\in \mathcal{S}}\alpha_{ijk'}},
\end{equation}
where $\sum_{k\in \mathcal{S}} \pi_{ijk}=1$. Using (\ref{1}), (\ref{2}), (\ref{10}), (\ref{11}), and (\ref{11a}), the average data rate that can be transmitted from node $i$ to node $j\in U_i$ for source node $k$ can be computed as
\begin{multline}\label{13}
  \hat{\mu}_{ijk}=\frac{\alpha_{ijk}}{\sum_{k'\in\mathcal{S}}\alpha_{ijk'}} \times\\
  \int_{\gamma=0}^{\infty}\frac{e^{-\frac{\gamma}{\bar{\gamma}_{ij}}}}{\bar{\gamma}_{ij}}
  \prod_{z\in U_i,z\neq j}(1-e^{-\frac{\beta_{ij}\gamma}{\beta_{iz}\bar{\gamma}_{ij}}})W\log_2(1+\gamma)d\gamma.
\end{multline}

Based on \eqref{7}, (\ref{8}), (\ref{9}) and (\ref{91}), the QoS constraint of source $k$ at node $i$ imposes a limit on the corresponding arrival rate as follows,
\begin{equation}\label{14}
  \rho_{ik}\leq \mu_{ik}-\frac{\ln\frac{1}{\epsilon^*_{ik}}}{D^*_{ik}}.
\end{equation}
Every node in the network restricts its transmitted data rate to each next hop node to avoid QoS violation at all subsequent hops on the path to the destination. These constraints determine the maximum data that node $i$ can forward for source $k\in \mathcal{S}$. Let $\hat{\rho}_{ijk}$ denote the maximum data rate for source $k$ that node $i$ can forward to next hop node $j$ such that all QoS constraints at $j$ and all subsequent hops to destination are satisfied. Also, let $\rho^*_{ik}$ denote the maximum arrival data rate of source $k$ to node $i$ such that QoS constraint at node $i$ and all subsequent hops to destination are satisfied. Therefore, we have
\begin{equation}\label{16b}
\rho^*_{ik}=\sum_{y\in V_i} \hat{\rho}_{yik},\,\, i\in {\cal{R}}_{L},
\end{equation}
where $V_i$ denotes the set of previous hop nodes of node $i$. At a last hop relay node $i\in \mathcal{R}_L$,
\begin{equation}\label{16}
\mu_{ikk}=\hat{\mu}_{ikk},\,\, i\in {\cal{R}}_{L}
\end{equation}
and
\begin{equation}\label{16a}
\rho^*_{ik}=\hat{\mu}_{ikk}-\frac{\ln\frac{1}{\epsilon^*_{ik}}}{D^*_{ik}}, \,\, i\in {\cal{R}}_{L}.
\end{equation}
In any other intermediate relay or source node $i\in\mathcal{S}\cup \mathcal{R}_1^{L-1}$, we have
\begin{equation}\label{17a}
 \mu_{ijk}=\min(\hat{\mu}_{ijk},\hat{\rho}_{ijk}), \,\, j\in U_i,
\end{equation}
and
\begin{multline}\label{17}
 \rho^*_{ik}= \sum_{j\in U_i} \min (\hat{\mu}_{ijk},\hat{\rho}_{ijk})-\frac{\ln\frac{1}{\epsilon^*_{ik}}}{D^*_{ik}}, \,\, i\in {\cal S}\cup {\cal R}_1^{L-1}.
\end{multline}
The right hand side of (\ref{17}) reflects the QoS constraint at node $i$ and all other subsequent hop nodes to destination. Therefore, to satisfy the QoS constraints, the arrival data rate of source $k$ is to be limited to
\begin{equation}\label{18}
  \rho_{k}=\rho^*_{kk},\,k\in{\cal S}.
\end{equation}

Using (\ref{16a}), (\ref{17}), and (\ref{18}), the problem (\ref{9}) may be rewritten as an unconstrained optimization problem for maximizing $F$ with respect to design parameters $\{\beta_{ij}\}$ and $\{\alpha_{ijk}\}$ as
\begin{multline}\label{19}
  \max_{\beta_{ij},\alpha_{ijk}} F=\sum_{k\in \mathcal{S}} f_k\rho^*_{kk},
  \\
  \rho^*_{ik}=\left\{
  \begin{array}{ll}
  \sum_{j\in U_i} \min (\hat{\mu}_{ijk},\hat{\rho}_{ijk})-\frac{\ln\frac{1}{\epsilon^*_{ik}}}{D^*_{ik}} & i\in \mathcal{R}_1^{L-1}  \\\\
  \hat{\mu}_{ikk}-\frac{\ln\frac{1}{\epsilon^*_{ik}}}{D^*{ik}}& i\in \mathcal{R}_{L}
  \end{array}\right.
\end{multline}

Every relay periodically (with period $T_1$) calculates the maximum arrival data rate that it can forward for each source using (\ref{16a}) or (\ref{17}), and restricts the maximum arrival data rate from each node in the previous hop accordingly. Specifically, node $i$ determines the maximum arrival data rate for source $k$ from node $y$ in the previous hop as follows,
\begin{equation}\label{24}
  \hat{\rho}_{yik}=\left\{
  \begin{array}{ll}
  \frac{\rho_{yik}}{\rho_{ik}} \rho^*_{ik} & \text{if }\rho_{ik}\neq 0\\\\
   \rho^*_{ik} & \text{if }\rho_{ik}= 0, y\in\mathcal{S}\\\\
  \frac{1}{M} \rho^*_{ik} & \text{if }\rho_{ik}= 0, y\not\in\mathcal{S}\\
  \end{array}\right.
\end{equation}
Also, source nodes set their average data rate based on (\ref{17}) and (\ref{18}), periodically with period $T_1$. We use gradient descent optimization to iteratively optimize $\{\beta_{ij}\}$ and $\{\alpha_{ijk}\}$ at the relays and source nodes. One important advantage of using this optimization method is that it can be implemented in a distributed way, which is appropriate for MANETs without any central control unit. Node $i$ updates $\alpha_{ijk}$ and $\beta_{ij}$ periodically with periods $T_1$ and $T_2$, respectively as follows
\begin{equation}\label{23}
  \alpha_{ijk}(n+1)=\alpha_{ijk}(n)+\theta_{\alpha} \frac{\partial F}{\partial \alpha_{ijk}},
\end{equation}
\begin{equation}\label{22}
  \beta_{ij}(m+1)=\beta_{ij}(m)+\theta_{\beta} \frac{\partial F}{\partial \beta_{ij}},
\end{equation}
where $n$ and $m$ and $\theta_{\alpha}$ and $\theta_{\beta}$ are the corresponding iteration numbers, and gradient decent step sizes. According to (\ref{18}) and (\ref{19}), we have
\begin{equation}\label{30}
   \frac{\partial F}{\partial \alpha_{ijk}}=\sum_{k\in\mathcal{S}}f_k\frac{\partial \rho_k}{\partial \alpha_{ijk}},
\end{equation}
\begin{equation}\label{30}
   \frac{\partial F}{\partial \beta_{ij}}=\sum_{k\in\mathcal{S}}f_k\frac{\partial \rho_k}{\partial \beta_{ij}}.
\end{equation}
According to (\ref{19}), at node $i$,
\begin{equation}\label{21}
  \frac{\partial \rho_k}{\partial \alpha_{ijk}}=\left\{
  \begin{array}{ll}
   \frac{\partial\rho_{ik}^*}{\partial \alpha_{ijk}}& \text{if  } \rho_{ik}=\rho_{ik}^*\\\\
  0 &\text{if  } \rho_{ik} < \rho_{ik}^*\\
  \end{array},
  \right.
\end{equation}
and
\begin{equation}\label{20}
  \frac{\partial \rho_k}{\partial \beta_{ij}}=\left\{
  \begin{array}{ll}
   \frac{\partial\rho_{ik}^*}{\partial \beta_{ij}}& \text{if  } \rho_{ik}=\rho_{ik}^*\\\\
  0 &\text{if  } \rho_{ik} < \rho_{ik}^*\\
  \end{array},
  \right.
\end{equation}
where $\partial\rho_{ik}^*/\partial \alpha_{ijk}$ and $\partial\rho_{ik}^*/\partial \beta_{ij}$ can be easily computed using (\ref{13}), (\ref{16a}), and (\ref{17}). Algorithm \ref{OCDR} summarizes the proposed OCDR scheme. We choose $T_2>\!>T_1>\!>T$ to ensure that a node has correctly estimated the arrival rate from previous hop nodes when updating its routing parameters.
%Also we set $T_2>\!>T_1$, thus routing parameters $\beta_{ij}$ are updated only when all the routing parameters $\alpha_{ijk}$ of nodes in the network and the arrival rates of source nodes are stable.

\begin{algorithm}
\caption{Distributed Opportunistic Cross-layer Dynamic Route Selection - OCDR} \label{OCDR}
\begin{itemize}
  \item Initially set $\rho_{k}=0$, $\beta_{ij}=1$, $\alpha_{ijk}=1$, $\hat{\rho}_{ijk}=0$, and $\rho^*_{ik}=0$ for all $k\in \mathcal{S}$, $i\in\mathcal{S}\cup\mathcal{R}_1^L$;\\

  \item In each time slot, node $i\in\mathcal{S}\cup\mathcal{R}_1^{L}$ chooses next link $j\in U_i$ for transmission based on (\ref{10}), and forwards data of source node $k$ with probability $\pi_{ijk}$ given in (\ref{11a});\\

   \item Node $i\in \mathcal{S}\cup \mathcal{R}_1^L$ updates $\alpha_{ijk}$ for all $j\in U_i$ and $k\in \mathcal{S}$ based on (\ref{23}) periodically with period $T_1$;\\

  \item Relay $i\in \mathcal{R}_1^L$ computes $\rho^*_{ik}$ using (\ref{16a}) or (\ref{17}), determines $\hat{\rho}_{yik}$ for all $y\in V_i$ using (\ref{24}), and informs $\hat{\rho}_{yik}$ to previous hop node $y\in V_i$ periodically with period $T_1$;\\

  \item Source node $k \in\mathcal{S}$ computes $\rho^*_{kk}$ using (\ref{17}) and sets the data rate $\rho_k=\rho_{kk}^*$ periodically with period $T_1$;\\

  \item Node $i\in \mathcal{S}\cup \mathcal{R}_1^L$ updates $\beta_{ij}$ for all $j\in U_i$ based on (\ref{22}) periodically with period $T_2$.\\

\end{itemize}
\end{algorithm}

\subsection{Time Division Cross-layer Dynamic Route Selection }
In the proposed time division cross-layer dynamic route selection (TCDR), each node assigns a fraction of time for transmission to next hop node based on the average quality of links. Let $\alpha_{ijk}'$ denote the fraction of time that node $i$ forwards data of source $k$ to node $j$ in the next hop. Thus, node $i$ forwards data of source $k$ to its next hop node $j\in U_i$ with probability
\begin{equation}\label{25}
  \pi'_{ijk}=\frac{\alpha_{ijk}'}{\sum_{k'\in\mathcal{S}}\sum_{j'\in U_i}\alpha_{ij'k'}'},
\end{equation}
where $\sum_{k\in\mathcal{S}}\sum_{j\in U_i}\pi'_{ijk}=1$. According to (\ref{1}), (\ref{2}), and (\ref{25}) the average data rate that can be transmitted from node $i$ to node $j\in U_i$ for source node $k$ can be computed as
\begin{equation}\label{26}
  \hat{\mu}_{ijk}=\frac{\alpha_{ijk}'}{\sum_{k'\in\mathcal{S}}\sum_{j'\in U_i}\alpha_{ij'k'}'} \times\\
  \int_{\gamma=0}^{\infty}\frac{e^{-\frac{\gamma}{\bar{\gamma}_{ij}}}}{\bar{\gamma}_{ij}}
  W\log_2(1+\gamma)d\gamma.
\end{equation}
The service rate at node $i$ for data of source $k$, $\mu_{ik}$ and the maximum arrival data rate of source node $k$ to node $i$ (such that all the QoS constraints at node $i$ and all subsequent hops to destination are satisfied),$\rho^*_{ik}$, are given by (\ref{16})-(\ref{17}). Using (\ref{16a}), (\ref{17}), and (\ref{18}), the problem (\ref{9}) can be rewritten as an unconstrained optimization problem for maximizing $F$ with respect to design parameters $\{\alpha'_{ijk}\}$ as
\begin{multline}\label{39}
  \max_{\alpha'_{ijk}} F=\sum_{k\in \mathcal{S}} f_k\rho^*_{kk},
  \\
  \rho^*_{ik}=\left\{
  \begin{array}{ll}
  \sum_{j\in U_i} \min (\hat{\mu}_{ijk},\hat{\rho}_{ijk})-\frac{\ln\frac{1}{\epsilon^*_{ik}}}{D^*_{ik}} & i\in \mathcal{R}_1^{L-1}  \\\\
  \hat{\mu}_{ikk}-\frac{\ln\frac{1}{\epsilon^*_{ik}}}{D^*{ik}}& i\in \mathcal{R}_{L}
  \end{array}\right.
\end{multline}

A relay node $i\in \mathcal{R}_1^l$ periodically (with period $T_1$) computes the maximum arrival data rate that it can forward for each source node using (\ref{16a}) or (\ref{17}), and restricts the maximum arrival data rate from each previous hop node $y\in V_i$ according to (\ref{24}). A source node $k\in \mathcal{S}$ sets their average data rate based on (\ref{17}) and (\ref{18}), periodically with period $T_1$. Similar to OCDR, in TCDR, we use gradient decent optimization to iteratively update $\alpha'_{ijk}$ at source nodes and relays. Node $i\in\mathcal{S}\cup\mathcal{R}_1^{L}$ updates $\alpha'_{ijk}$ for all $j\in U_i$ and $k\in \mathcal{S}$, periodically with period $T_1$, as follows
\begin{equation}\label{63}
  \alpha'_{ijk}(n+1)=\alpha'_{ijk}(n)+\theta_{\alpha'} \frac{\partial F}{\partial \alpha'_{ijk}},
\end{equation}
where $n$ is the iteration number and $\theta_{\alpha'}$ is the gradient decent step size. According to (\ref{18}) and (\ref{39}), we have
\begin{equation}\label{50}
   \frac{\partial F}{\partial \alpha'_{ijk}}=\sum_{k\in\mathcal{S}}f_k\frac{\partial \rho_k}{\partial \alpha'_{ijk}}.
\end{equation}
Based on (\ref{39}), at node $i$,
\begin{equation}\label{52}
  \frac{\partial \rho_k}{\partial \alpha'_{ijk}}=\left\{
  \begin{array}{ll}
   \frac{\partial\rho_{ik}^*}{\partial \alpha'_{ijk}}& \text{if  } \rho_{ik}=\rho_{ik}^*\\\\
  0 &\text{if  } \rho_{ik} < \rho_{ik}^*\\
  \end{array},
  \right.
\end{equation}
where $\partial\rho_{ik}^*/\partial \alpha'_{ijk}$ can be easily computed using (\ref{16a}), (\ref{17}), and (\ref{26}). Algorithm \ref{TCDR} summarizes the proposed TCDR scheme. We choose $T_1>\!>T$ to ensure that a node has enough time to estimate its arrival rate from previous hop nodes before updating its routing parameters.
%Also we set $T_2>\!>T_1$, thus routing parameters $\beta_{ij}$ are updated only when all the routing parameters $\alpha_{ijk}$ of nodes in the network and the arrival rates of source nodes are stable.
The simulation results show that OCDR provides a higher performance compared to TCDR. However, TCDR does not need access to instantaneous channel SNRs and changes the routes only periodically. Therefore, TCDR has a lower implementation complexity.\\

\begin{algorithm}
\caption{Distributed Time Division Cross-layer Dynamic Route Selection - TCDR} \label{TCDR}
\begin{itemize}
  \item Initially set $\rho_{k}=0$, $\alpha'_{ijk}=1$, $\hat{\rho}_{ijk}=0$, and $\rho^*_{ik}=0$ for all $k\in \mathcal{S}$, $i\in\mathcal{S}\cup\mathcal{R}_1^L$;\\

  \item In each time slot, node $i\in\mathcal{S}\cup\mathcal{R}_1^{L}$ chooses next link $j\in U_i$ for transmission and forwards data of source node $k$ with probability $\pi'_{ijk}$ given in (\ref{25});\\

   \item Node $i\in \mathcal{S}\cup \mathcal{R}_1^L$ updates $\alpha'_{ijk}$ for all $j\in U_i$ and $k\in \mathcal{S}$ based on (\ref{63}) periodically with period $T_1$;\\

  \item Relay $i\in \mathcal{R}_1^L$ computes $\rho^*_{ik}$ using (\ref{16a}) or (\ref{17}), determines $\hat{\rho}_{yik}$ for all $y\in V_i$ using (\ref{24}), and informs $\hat{\rho}_{yik}$ to previous hop node $y\in V_i$, periodically with period $T_1$;\\

  \item Source node $k \in\mathcal{S}$ computes $\rho^*_{kk}$ using (\ref{17}) and sets the data rate $\rho_k=\rho_{kk}^*$, periodically with period $T_1$.\\

\end{itemize}
\end{algorithm}

\section{Performance Evaluation}
\begin{figure}
\centering
\subfigure[two hop ($L=1$)]{\includegraphics[width=3.4in, trim=0cm 7.3cm 0cm 6cm, clip=true]{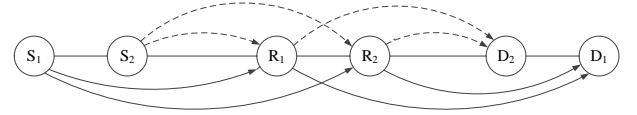}\label{LN2}}
\subfigure[three hop ($L=2$)]{\includegraphics[width=3.4in, trim=0cm 7.3cm 0cm 6cm, clip=true]{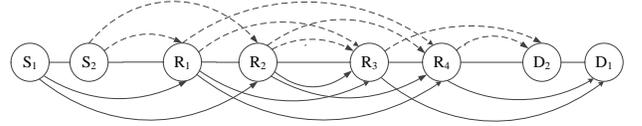} \label{LN3}}
\caption{Linear network  model ($M=N=2$). } \label{LN}
\end{figure}

We consider multihop transmission in a one dimensional network. The linear network model for $N=M=2$ and $L=1,2$ is illustrated in Figure \ref{LN}. We use standard path loss model to model signal attenuation over each link. Thus the average channel gain between nodes $i$ and $j$ is $\bar{h}_{ij}=cd_{ij}^{-\delta}$, where $c$ is a constant, $d_{ij}$ is the distance between nodes $i$ and $j$, and $\delta$ is the path loss exponent. We set ${cP\over{N_0W}}=1$, $\delta=3$, and $W=1\text{MHz}$. The maximum tolerable delay in each queue, $D_{ik}^*$, and the packet loss rate threshold in node $i$ for data of source $k$, $\epsilon_{ik}^*$, are set at $0.1\text{ms}$ and $10^{-6}$, respectively. The benchmark for comparison is the maximum possible weighted sum-rate performance, by static assignment of one relay in each hop to each source destination pair. This is obtained by examining all possible relay assignment scenarios, while the QoS constraints are satisfied.\\

\begin{figure*}
\centering
\subfigure[OCDR]{\includegraphics[width=2.31in, trim=.8cm 3.4cm .95cm .95cm, clip=true]{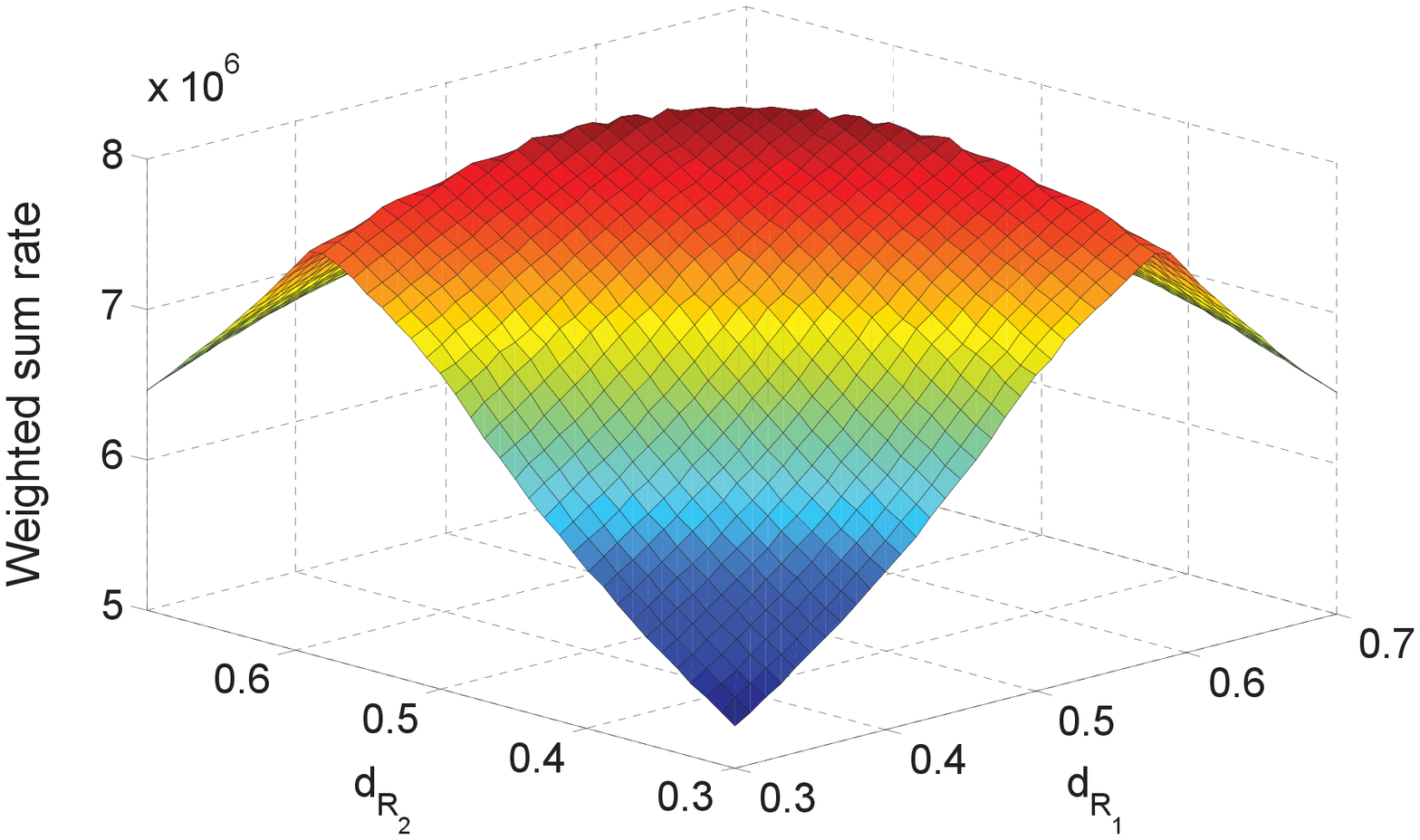}\label{fig3}}
\subfigure[TCDR]{\includegraphics[width=2.31in, trim=.8cm 3.4cm .95cm .95cm, clip=true]{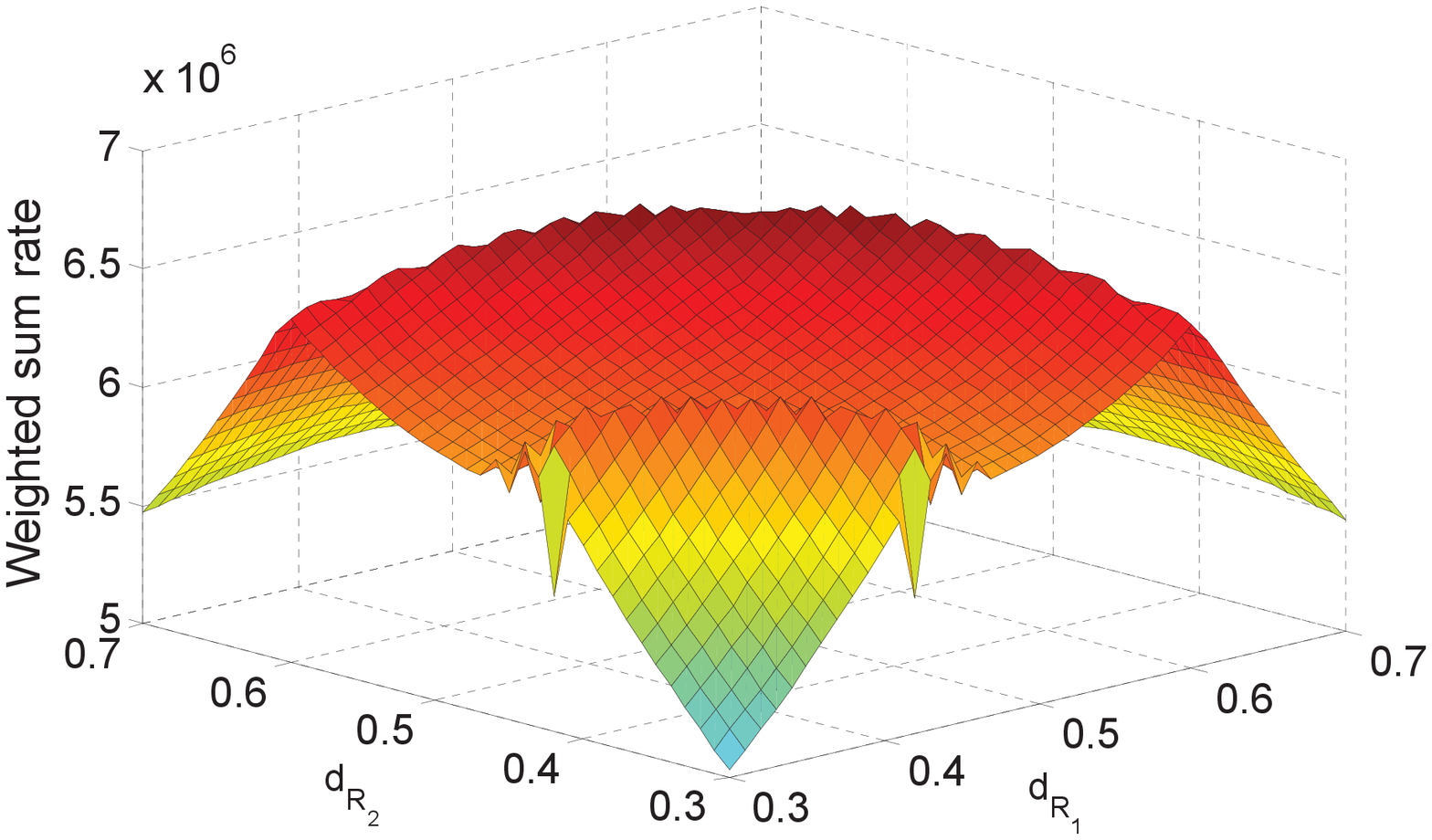} \label{fig4}}
\subfigure[Benchmark]{\includegraphics[width=2.31in, trim=.8cm 3.4cm .95cm .95cm, clip=true]{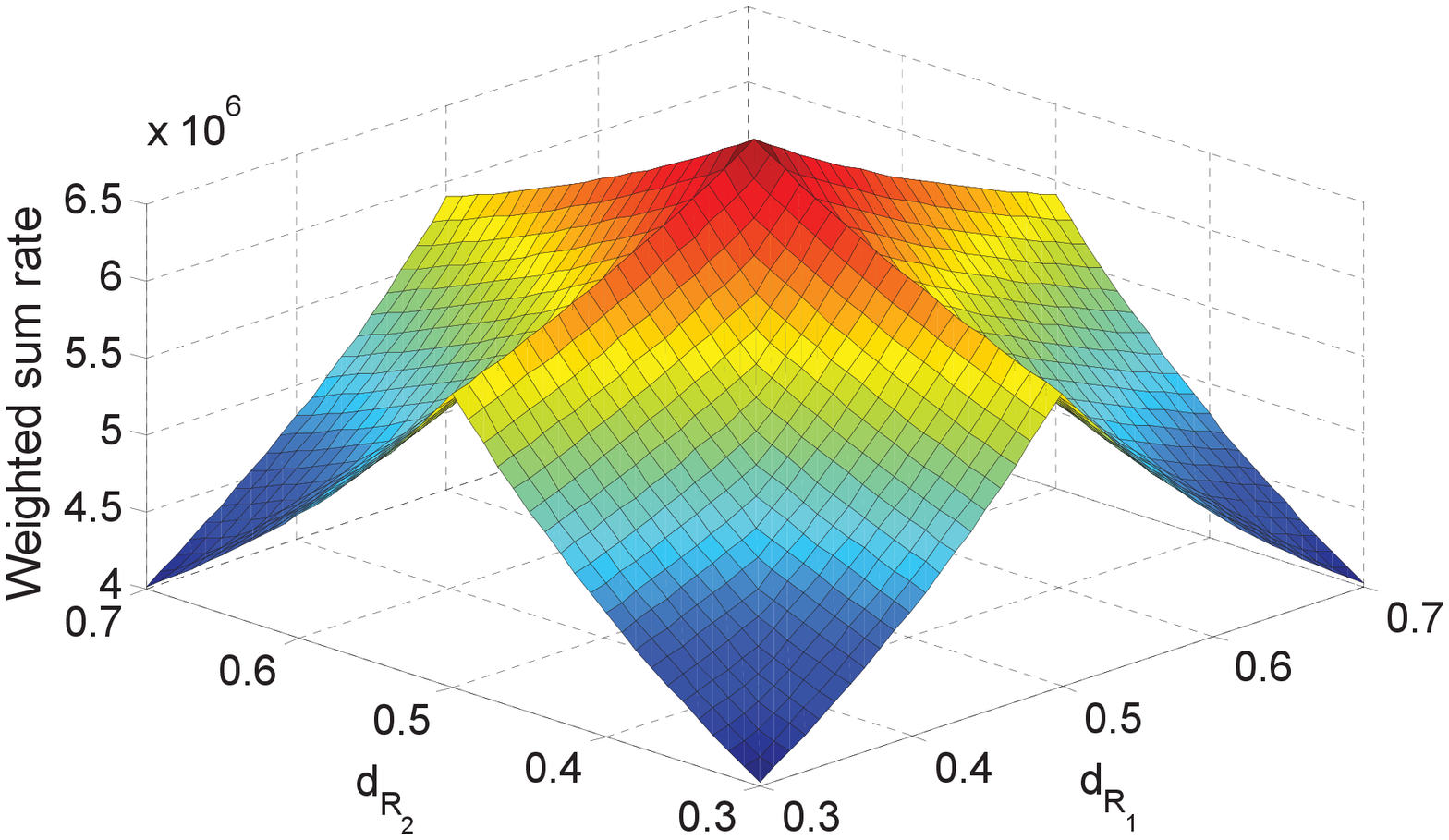} \label{fig5}}
\caption{Sum-rate performance of OCDR, TCDR, and benchmark in a two-hop network ($L=1$).} \label{2P}
\end{figure*}

\begin{figure*}
\centering
\subfigure[OCDR vs benchmark]{\includegraphics[width=2.31in, trim=0.8cm 3.4cm .95cm .95cm, clip=true]{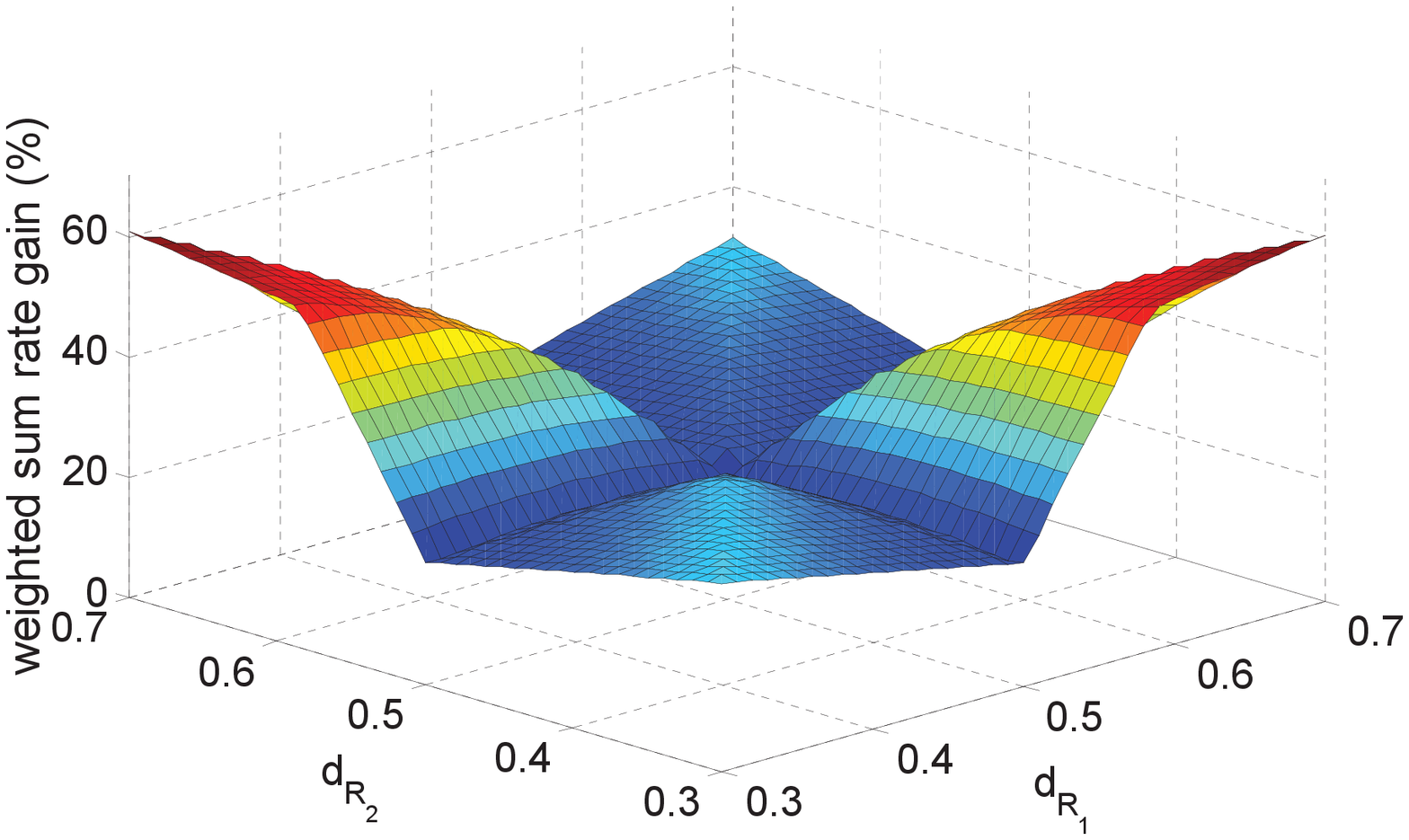}\label{fig6}}
\subfigure[TCDR vs benchmark]{\includegraphics[width=2.31in, trim=0.8cm 3.4cm .95cm .95cm, clip=true]{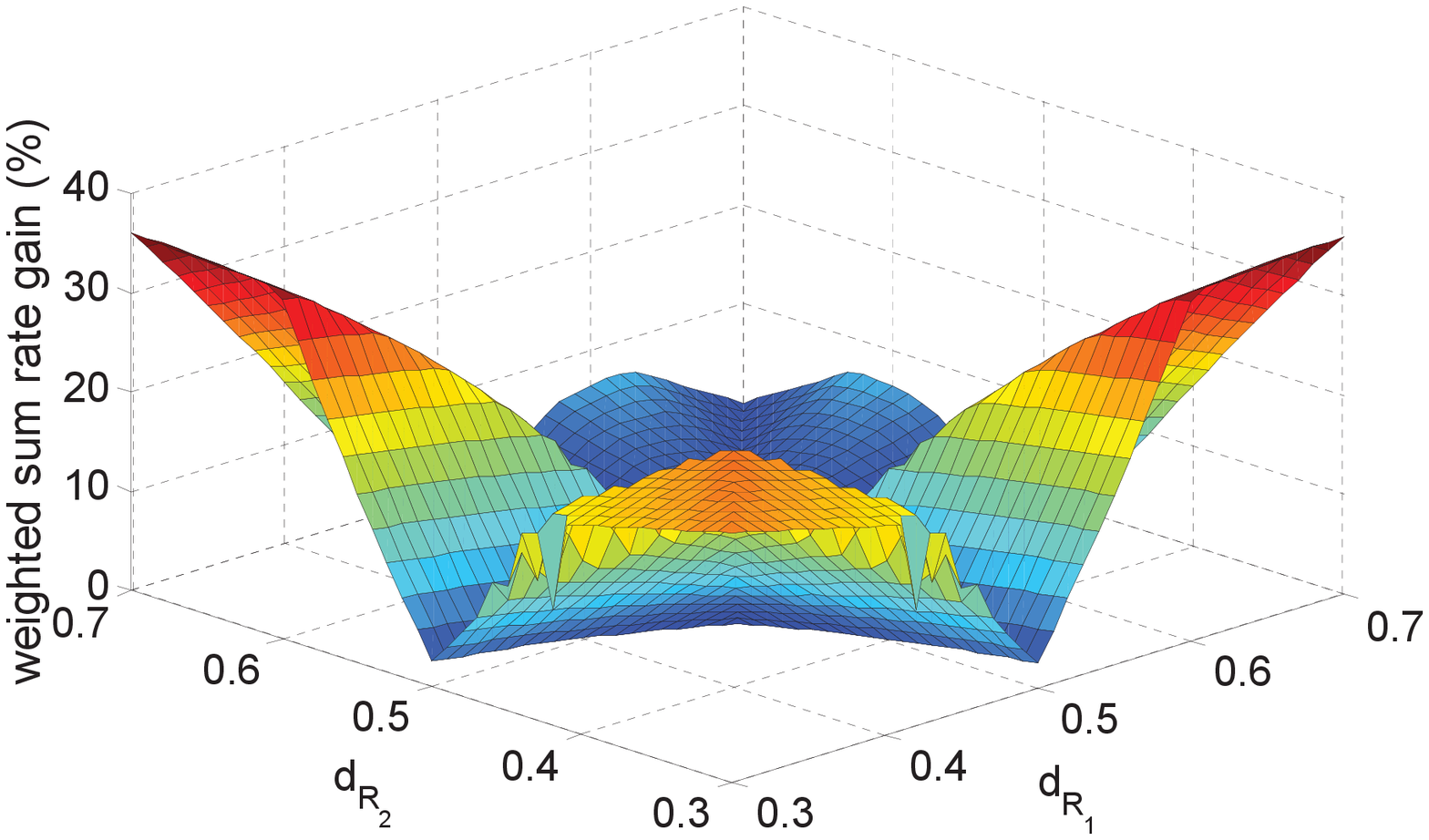} \label{fig7}}
\subfigure[OCR vs TCDR]{\includegraphics[width=2.31in, trim=0.8cm 3.4cm .95cm .95cm, clip=true]{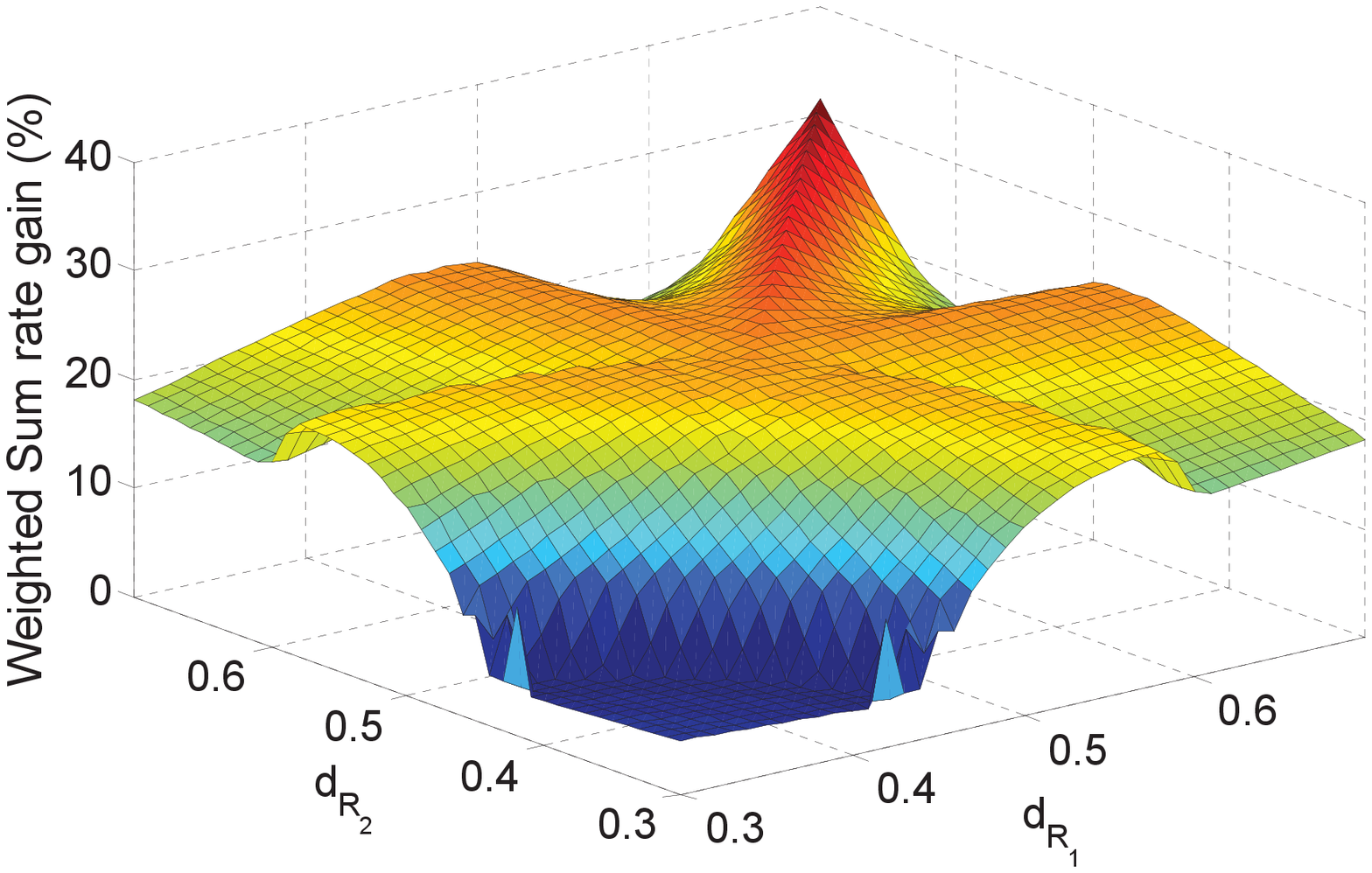} \label{fig8}}
\caption{Sum-rate performance gain ($\%$)  of OCDR, TCDR, and benchmark in a two-hop network ($L=1$).} \label{2G}
\end{figure*}

\begin{figure*}
\centering
\subfigure[OCDR]{\includegraphics[width=2.31in, trim=0.8cm 3.4cm .95cm .95cm, clip=true]{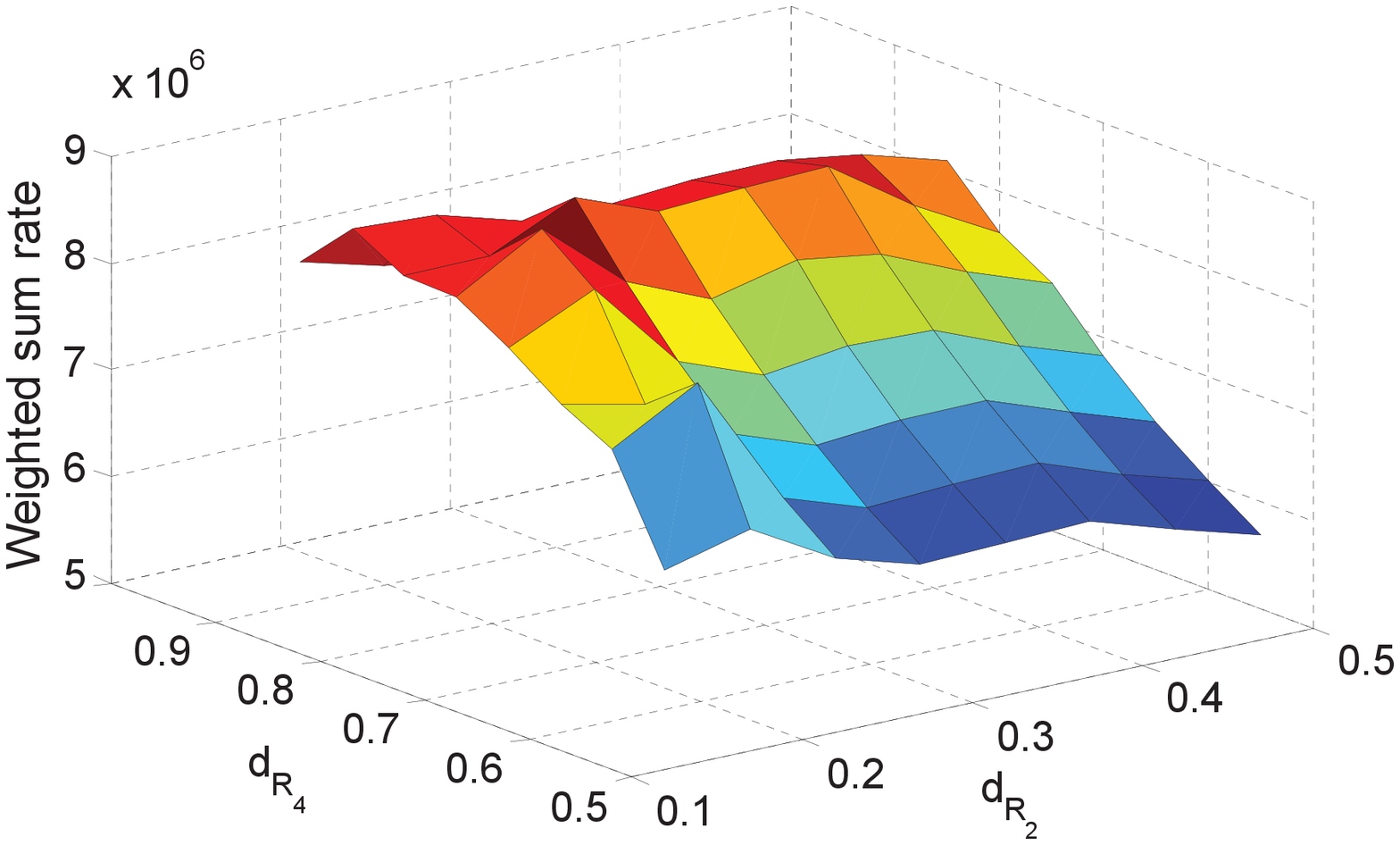}\label{fig10}}
\subfigure[TCDR]{\includegraphics[width=2.31in, trim=0.8cm 3.4cm .95cm .95cm, clip=true]{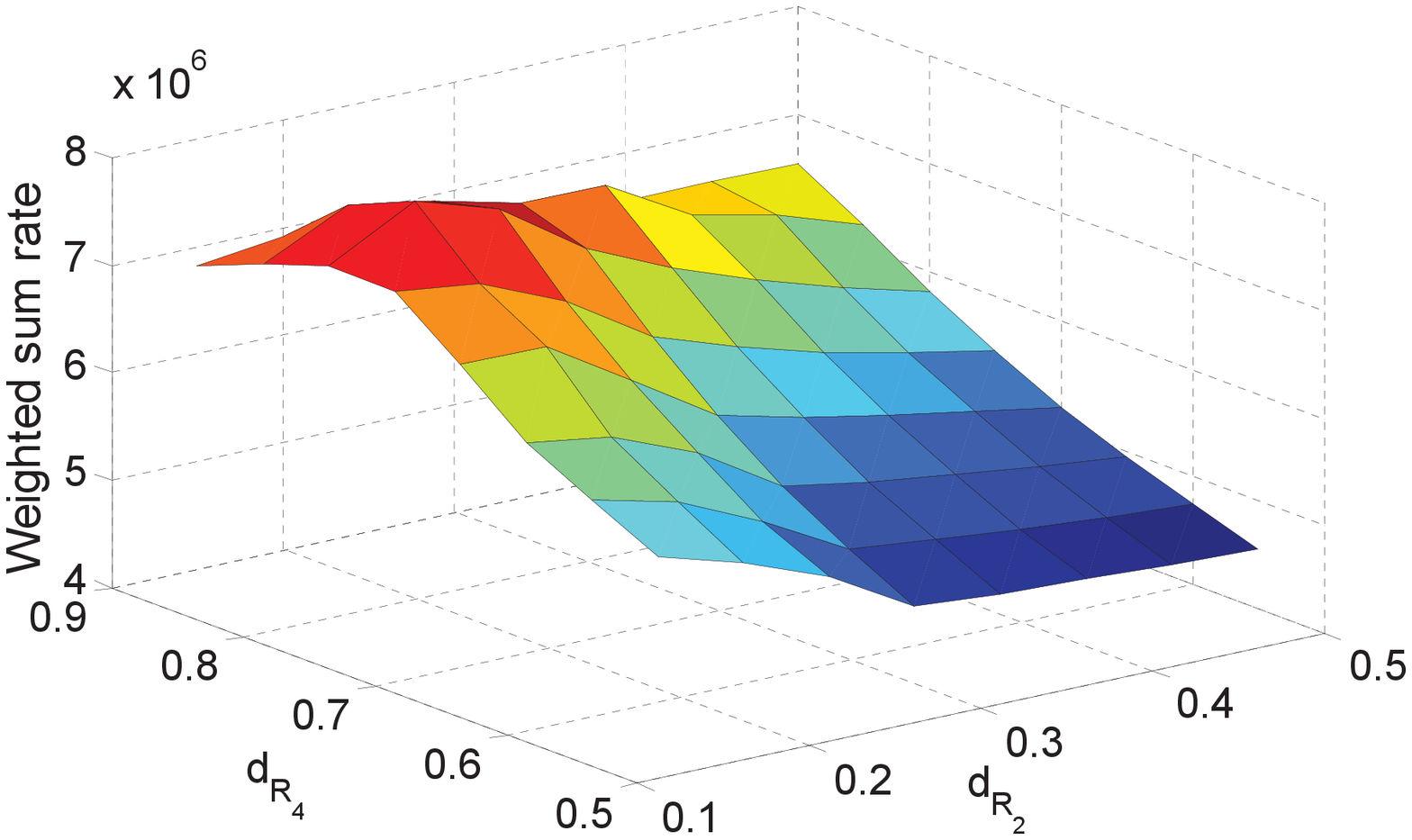} \label{fig11}}
\subfigure[Benchmark]{\includegraphics[width=2.31in, trim=0.8cm 3.4cm .95cm .95cm, clip=true]{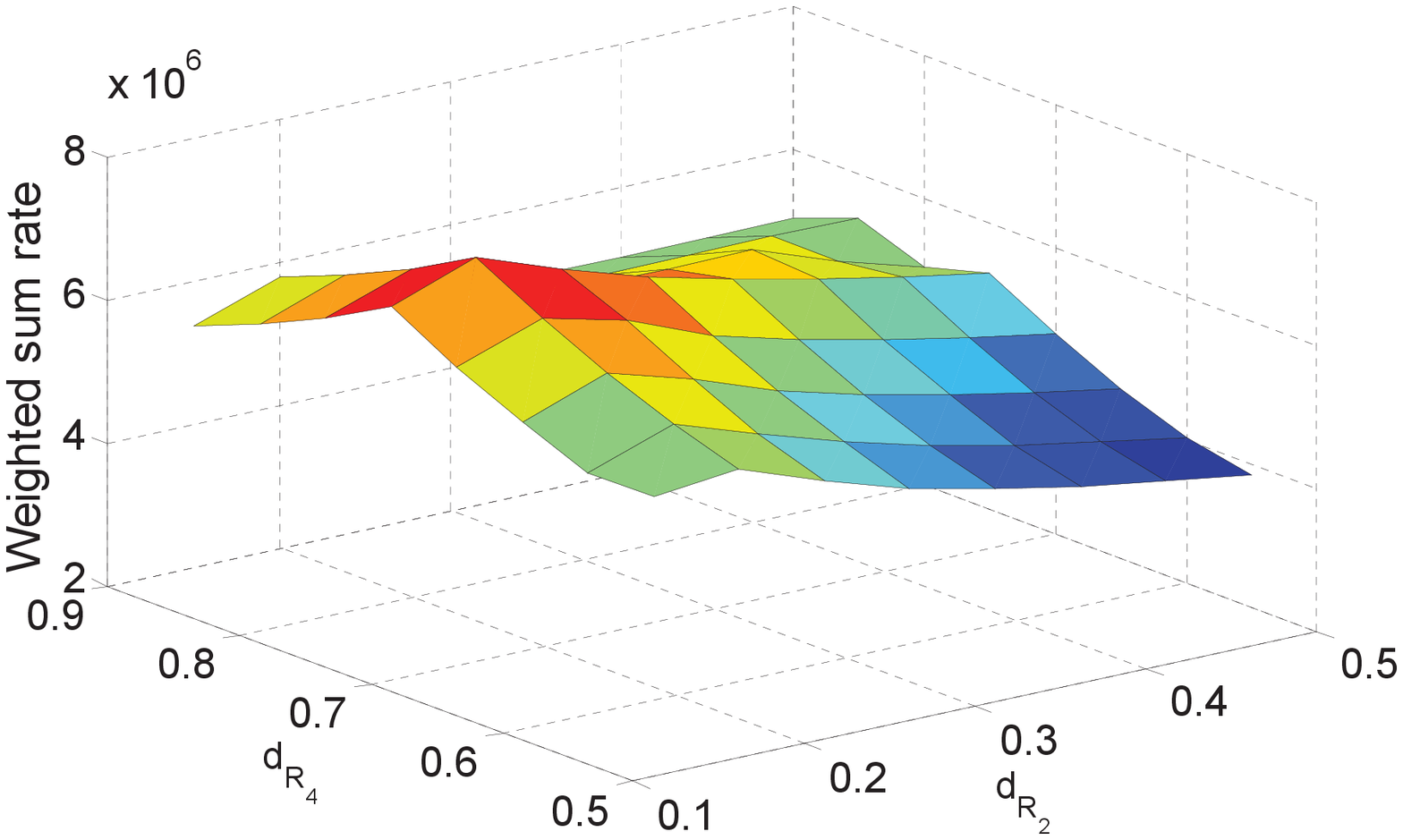} \label{fig12}}
\caption{Sum-rate performance of OCDR, TCDR, and benchmark in a three-hop network ($L=2$).} \label{3P}
\end{figure*}

\begin{figure*}
\centering
\subfigure[OCDR vs benchmark]{\includegraphics[width=2.31in, trim=0.8cm 3.4cm .95cm .95cm, clip=true]{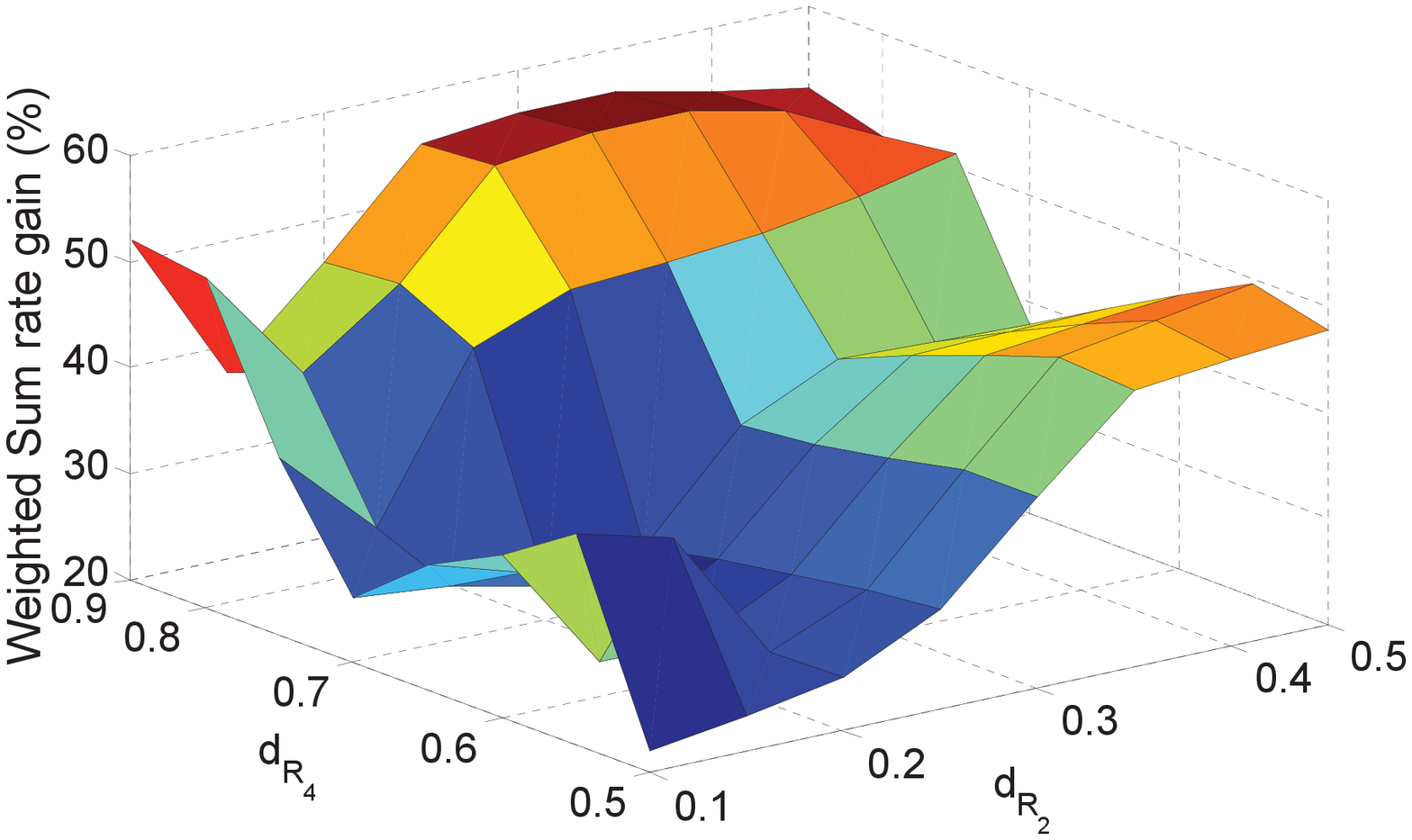}\label{fig13}}
\subfigure[TCDR vs benchmark]{\includegraphics[width=2.31in, trim=0.8cm 3.4cm .95cm .95cm, clip=true]{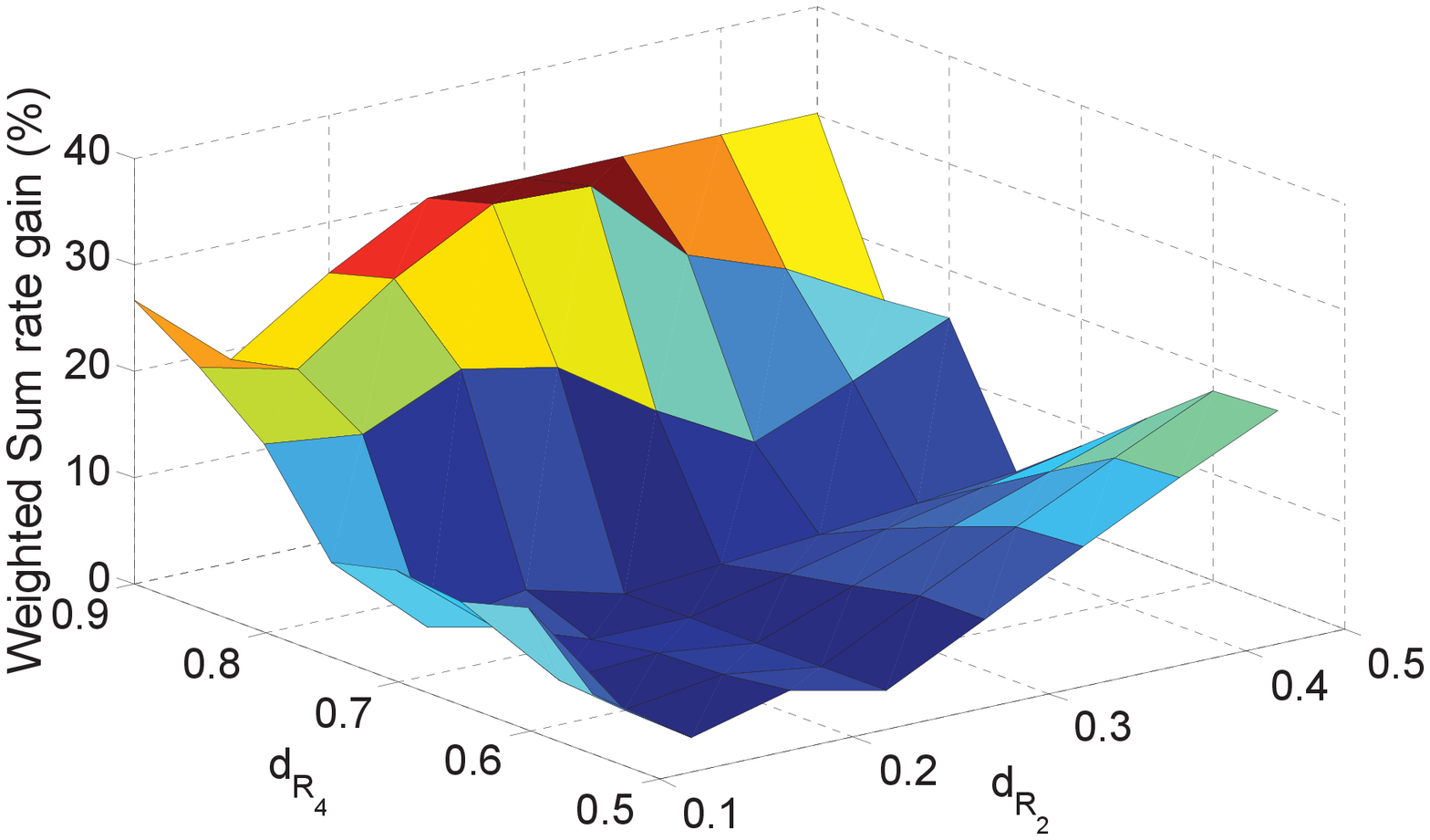} \label{fig14}}
\subfigure[OCDR vs TCDR]{\includegraphics[width=2.31in, trim=0.8cm 3.4cm .95cm .95cm, clip=true]{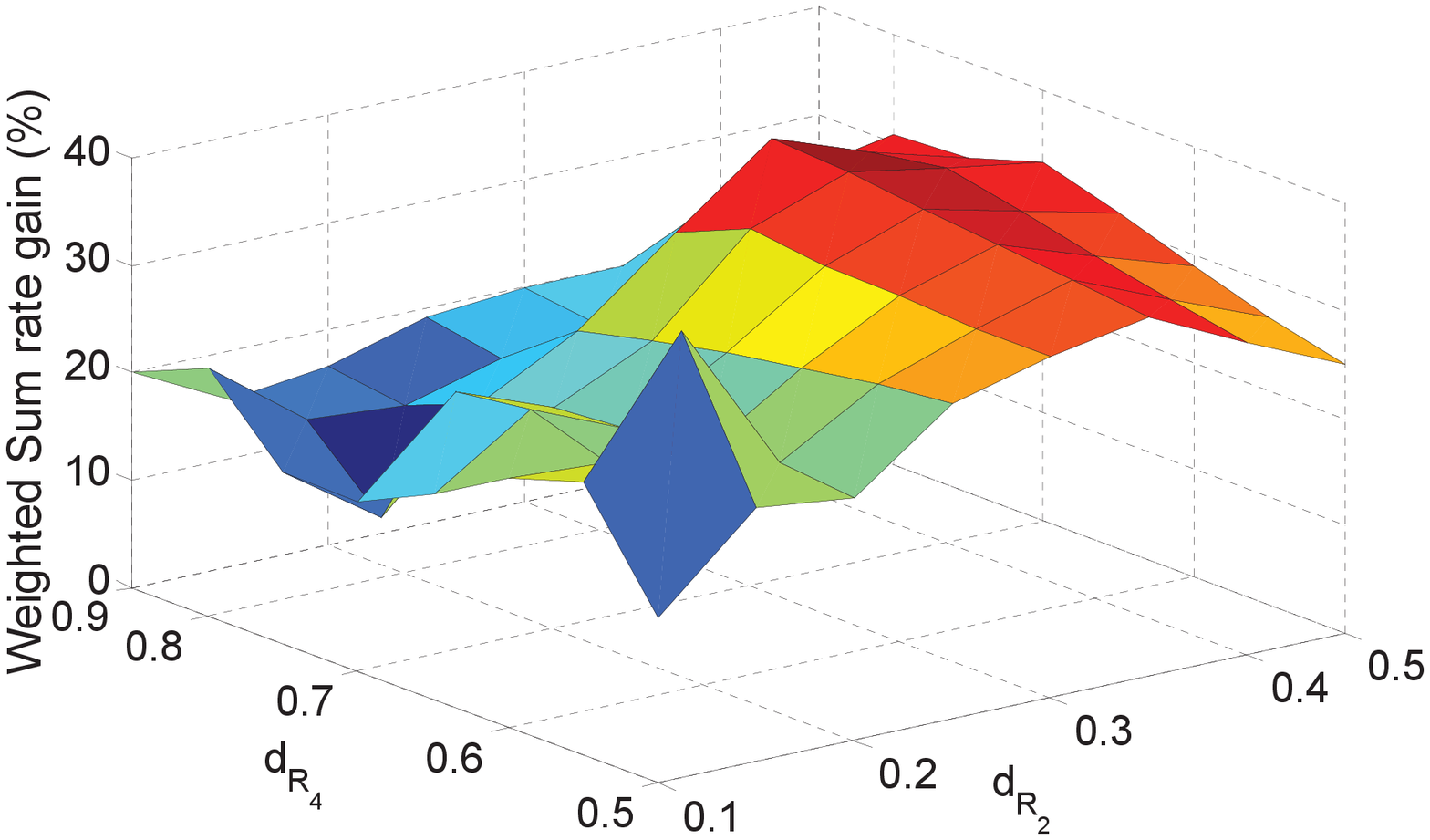} \label{fig14N2}}
\caption{Sum-rate performance gain ($\%$)  of OCDR, TCDR, and benchmark in a three-hop network ($L=2$).} \label{3G}
\end{figure*}

\begin{figure}
\centering
\subfigure[Average data rate of source destination pairs]{\includegraphics[width=3.4in, trim=0cm 1.3cm 1cm 0cm, clip=true]{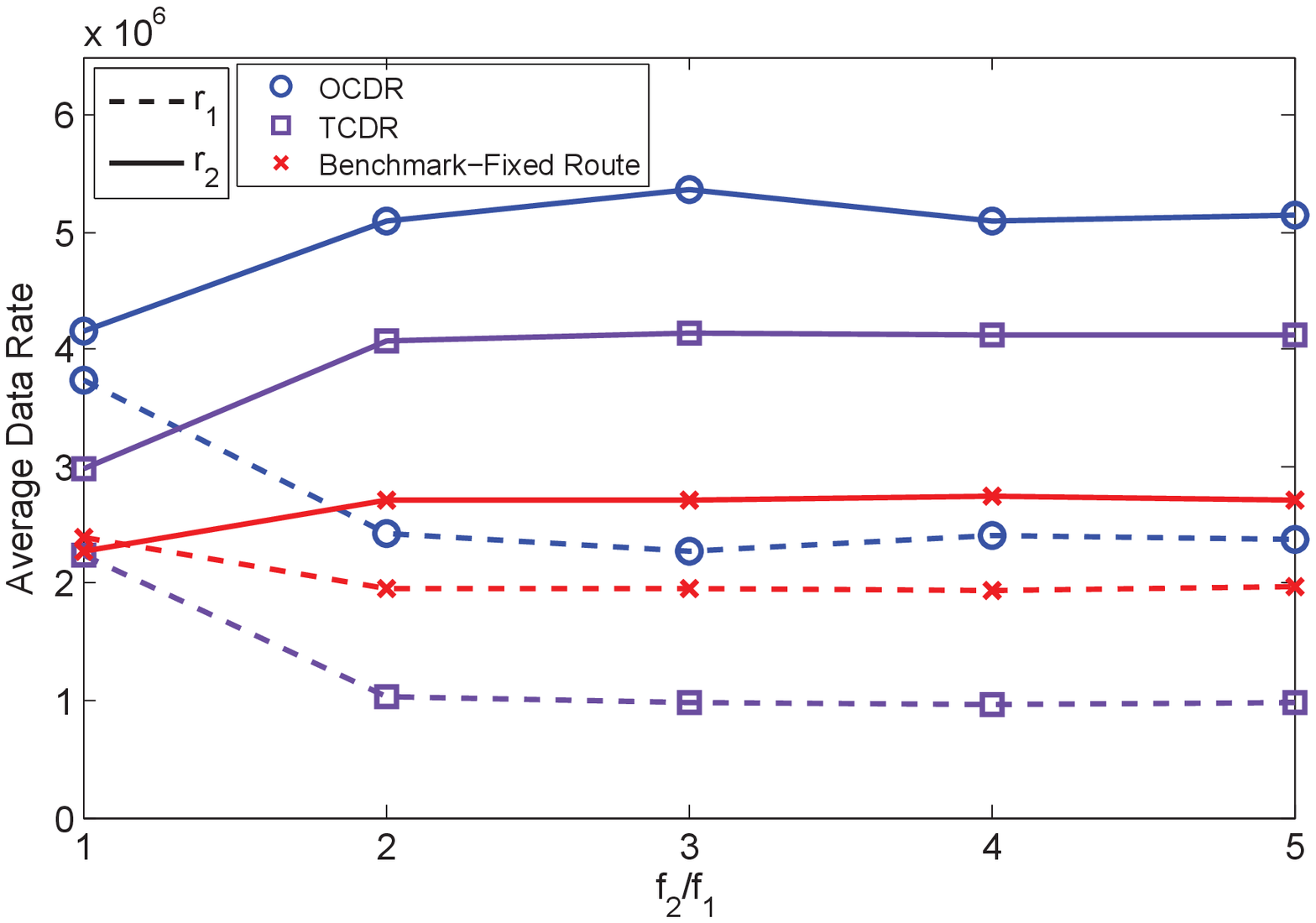}\label{fig55.eps}}
\subfigure[Average weighted sum rate]{\includegraphics[width=3.4in, trim=0cm 1.3cm 1cm 0cm, clip=true]{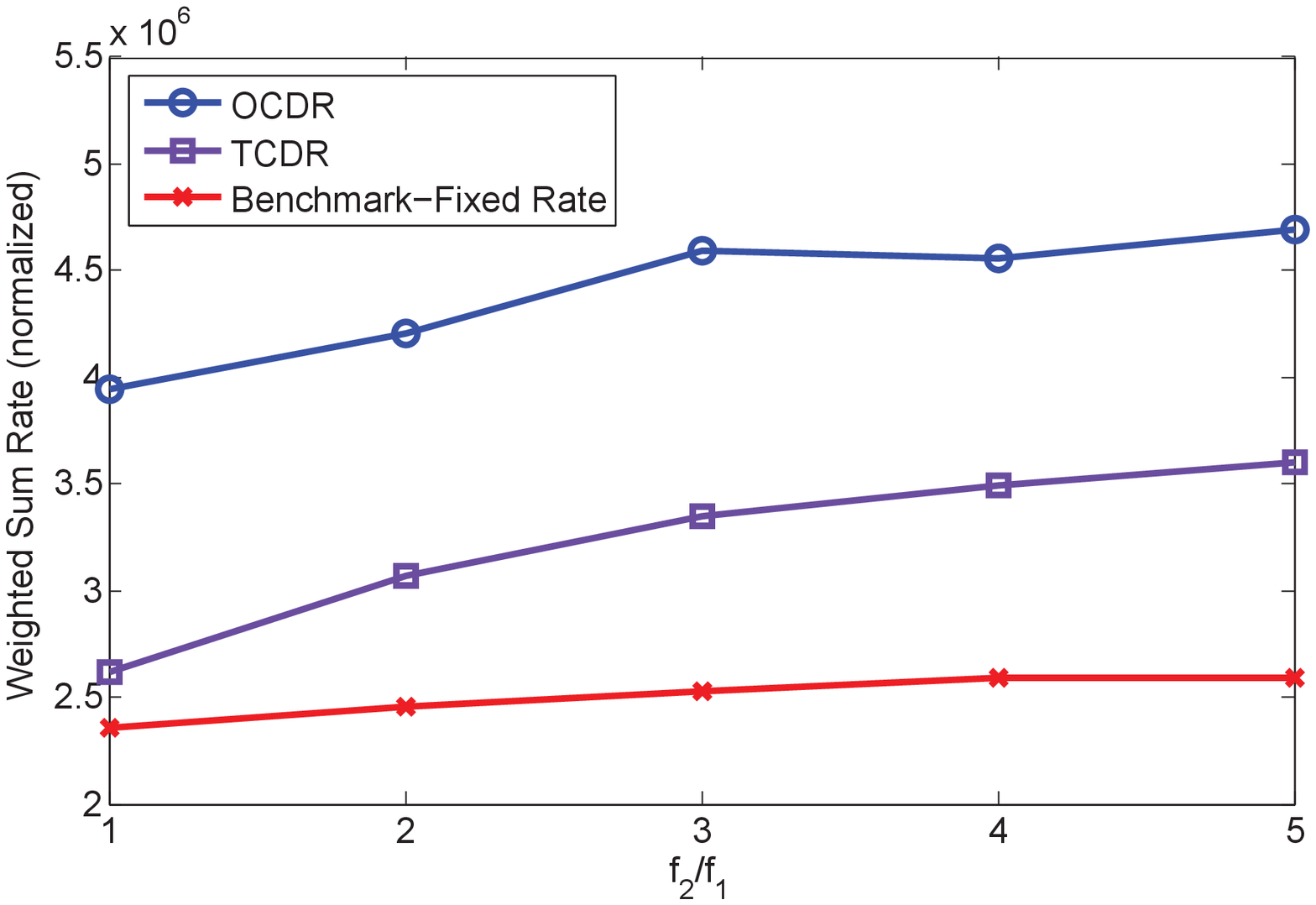}\label{fig66}}
\caption{Average performance of OCDR, TCDR and benchmark for 100 random relay positions ($L=2$) vs assigned weights.} \label{fig5566}
\end{figure}

\subsection{Two hop network}
Consider two-hop transmission in a network consisting of two source-destination pairs and two intermediate relays ($K=M=2, L=1$). Figure \ref{LN2} shows the network model. The first source and destination pair are positioned at $d_{S_1}=0$ and $d_{D_1}=1$, respectively. The second source and destination pair are positioned at $d_{S_2}=0.2$ and $d_{D_2}=0.8$, respectively. Relays are positioned at $d_{R_1}$ and $d_{R_2}$. The relays positions vary such that $0.3\leq d_{R_1}\leq 0.7$ and $0.3\leq d_{R_2}\leq 0.7$. We aim to maximize the sum-rate, i.e., we set $f_1=f_2=1$.\\

Figure \ref{2P} shows the sum-rate performance as a function of the relays positions for OCDR, TCDR, and the benchmark scheme. According to Figure \ref{2P}, the sum-rate is maximized in each of the schemes (OCDR, TCDR, or benchmark) when relay nodes are positioned at points $d_{R_1}=0.5$ and $d_{R_2}=0.5$, which corresponds to the situation that the channel quality of first and second hops are identical for each source-destination pair. While fixed relay assignment (benchmark) provides very poor performance when channel quality of links are diverse, the proposed OCDR and TCDR schemes provide significantly higher performance by taking advantage of transmitting data through dynamic routes. Figures \ref{fig6} and \ref{fig7} illustrate the sum-rate performance gains of OCDR and TCDR with respect to the benchmark, respectively. On average, OCDR and TCDR provide $32\%$ and $11\%$ performance gain, however, this gain may reach $60\%$ and $40\%$ depending on the relays positions. The sum-rate performance gain of OCDR with respect to TCDR is shown in Figure \ref{fig8}. The OCDR when compared to TCDR enhances the weighted sum-rate performance by up to $32\%$ depending on the relays positions. This comes at the cost of instantaneous channel state information and higher implementation complexity. Indeed, the proposed TCDR scheme only requires the average quality of links and alternates the data transmission path periodically.\\

\subsection{Three hop network}
In this Subsection, we evaluate the performance of the proposed schemes for three-hop transmission in a linear network consisting of two source-destination pairs and four intermediate relays ($K=M=2,L=2$). The network model is illustrated in Figure \ref{LN3}. The source nodes are placed at $d_{S_1}=d_{S_2}=0$ and destination nodes are placed at $d_{D_1}=d_{D_2}=1$. First hop relays are $R_1$ and $R_2$ and second hop relays are $R_3$ and $R_4$, which are positioned at $d_{R_1}$, $d_{R_2}$, $d_{R_3}$, and $d_{R_4}$, respectively.\\

Figure \ref{3P} shows the sum-rate performance ($f_1=f_2=1$) versus relays position for OCDR, TCDR, and the benchmark scheme, when $d_{R_1}=0.2$, $0.1\leq d_{R_2}\leq 0.5$, $d_{R_3}=0.7$, and $0.5\leq d_{R_4} \leq 0.9$. The maximum sum-rate in each scheme corresponds to the situation that the link quality of hops are identical, i.e., $d_{R_2}=0.1$ and $d_{R_4}=0.7$. When the quality of links are diverse, the static relay assignment (benchmark) provides a very poor performance. However, the proposed OCDR and TCDR schemes provide significantly higher performance by directing data through multiple dynamic routes to destinations. Figures \ref{fig13} and \ref{fig14} illustrate the sum-rate performance gains of OCDR and TCDR with respect to the benchmark. Also, the sum-rate performance gain of OCDR with respect to TCDR is shown in Figure \ref{fig14N2}. According to Figure \ref{3G}, the performance gain of proposed OCDR and TCDR schemes with respect to the benchmark may reach $55\%$ and $30\%$ depending on relays positions.\\

\subsection{Effect of weights}
In this Subsection, we evaluate the impact of weights on the average data rate of each source destination pair in a three-hop linear network as illustrated in Figure \ref{LN3}. The source nodes are placed at $d_{S_1}=d_{S_2}=0$ and destination nodes are placed at $d_{D_1}=d_{D_2}=1$. The four relays are randomly positioned in the range $[0.1,0.9]$ and in each realization, the two relays closer to source nodes are selected as the first hop relays, $R_1$ and $R_2$, and the other two relays are selected as the second hop relays, $R_3$ and $R_4$. Figures \ref{fig55} and \ref{fig66} depict respectively, the average rates, $r_1$ and $r_2$, of each source-destination pair and the average weighted sum rate (normalized to the sum of weights), for one hundred random relays positions as a function of the weights assigned to source destination pairs ($f_1$ and $f_2$). One sees that as $f_2/f_1$ increases, the data rate of second source destination pair increases in OCDR, TCDR and the benchmark. The proposed OCDR and TCDR schemes provide significantly higher average sum-rate and better performance differentiation between the users in comparison to the benchmark.\\

\section{Conclusions}
In this paper, we present two new cross-layer dynamic route selection schemes for multiuser multihop transmission in wireless ad-hoc networks. The proposed schemes set up the routes between source-destination pairs in a cross-layer optimized manner, which takes into account the buffer status of network nodes and the instantaneous condition of fading channels or the average link qualities. The proposed schemes are distributed. Nodes exchange routing information with single hop neighbors and iteratively adjust the routing parameters to optimize the network performance. The simulation results show the effectiveness of our proposed dynamic route selection schemes in comparison with conventional fixed routing.\\

\appendices

\ifCLASSOPTIONcaptionsoff
 % \newpage
\fi

\bibliographystyle{IEEEtran}
\bibliography{myrefMSC}
%\bibliographystyle{IEEEtran}
%\bibliography{myref}

%\begin{thebibliography}{1}
%\end{thebibliography}
%
%\begin{IEEEbiography}{Kamal Rahimi Malekshan}
%The biography text here.
%\end{IEEEbiography}
%

\end{document}